\documentclass[twocolumn,prl,showpacs,preprintnumbers,amsmath,amssymb,floatfix]{revtex4}

\usepackage{graphicx}
\usepackage{dcolumn}
\usepackage{bm}
\usepackage{verbatim}
\usepackage{tabularx}

\def\beq{\begin{equation}}
\def\eeq{\end{equation}}
\def\bea{\begin{eqnarray}}
\def\eea{\end{eqnarray}}

\newcommand{\Fig}[1]{Fig.~\ref{#1}}

\def\Zp{Z^\prime}
\def\Afb{A_{FB}^{t}}
\def\Mtt{M_{t\bar{t}}}

\begin{document}

\preprint{MCTP/09-35, CERN-PH-TH/2009-128, IPMU09-0089}

\title{Top quark forward-backward asymmetry from new $t$-channel physics}

\author{Sunghoon Jung$^1$, Hitoshi Murayama$^{2,3,4}$, Aaron Pierce$^1$, James D. Wells$^{1,5}$}
\affiliation{
$^1$ Michigan Center for Theoretical Physics,
University of Michigan, Ann Arbor, MI 48109, USA \\
$^2$ Department of Physics, University of California, Berkeley, CA 94720, USA\\
$^3$ Theoretical Physics Group, LBNL, Berkeley, CA 94720, USA\\
$^4$ IPMU, University of Tokyo, 5-1-5 Kashiwa-no-ha, Kashiwa, Japan 277-8568\\
$^5$ CERN Theoretical Physics (PH-TH), CH-1211 Geneva 23, Switzerland
}

\date{\today}

\begin{abstract}
Motivated by recent measurements of the top quark forward-backward asymmetry at the Tevatron, we study how $t$-channel new physics can contribute to a large value.  We concentrate on a theory with an abelian gauge boson possessing flavor changing couplings between up and top quarks, but satisfies flavor physics constraints. Collider constraints are strong, but can be consistent with the aid of small flavor diagonal couplings. We find that $M_{\Zp} \approx 160\,$GeV can yield a total lab-frame asymmetry of $\sim 18\%$ without being in conflict with other observables. There are implications for future collider searches, including exotic top quark decays, like-sign top quark production, and detailed measurements of the top production cross section.  An alternate model with a gauged non-Abelian flavor symmetry would have similar phenomenology, but lacks the like-sign top signal.
\end{abstract}

\maketitle


\textbf{Introduction.}
 The most recent measurement of the top quark forward-backward asymmetry is from the CDF experiment, which obtains $\Afb = 19.3 \pm 6.9 \%$ with 3.2 fb$^{-1}$ of data \cite{cdf:afb}. The Standard Model (SM) prediction \cite{Kuhn:1998jr,Kuhn:1998kw,Bowen:2005ap,Almeida:2008ug} is dominated by ${\mathcal O}(\alpha_{S}^{3})$ QCD interference effects and is $5\%$ in the lab frame. At present, this discrepancy is less than $3\sigma$. However, it is interesting to ask whether such a large central value can be explained, especially once one accounts for the other Tevatron measurements of top quark properties, all consistent with the SM. It is intriguing that past measurements at CDF and D0 have yielded consistently large  asymmetry values~\cite{Aaltonen:2008hc,abazov:2007qb}.

Many models of new physics impact $\Afb$, but it is difficult to produce a large positive asymmetry. The most constrained idea is perhaps axigluons, which interfere with QCD and induce large negative asymmetries~\cite{Ferrario:2008wm,Antunano:2007da,Choudhury:2007ux}. Kaluza-Klein excitations of the gluon in warped AdS space may produce positive asymmetries~\cite{Djouadi:2009nb}. 

\textbf{Model.}
Our model consists of a new vector boson ($\Zp$) associated with an abelian gauge symmetry $U(1)_{\Zp}$ with flavor off-diagonal couplings 
${\mathcal L} \ni g_{X} \Zp_{\mu} \bar{u} \gamma^{\mu} P_{R} t +h.c.$ This can generate $\Afb$ through $t$-channel exchange of $Z^{\prime}$, $u\bar u \to t\bar t$.
We also allow a small flavor-diagonal coupling to up-type quarks ${\mathcal L} \ni \epsilon_{U} g_{X} \Zp_{\mu} \bar{u}_i \gamma^{\mu} P_{R} u_i$, with $\epsilon_{U} < 1$ and generation index $i$.  If no diagonal coupling for the $\Zp$ exists ($\epsilon_U=0$), it is forced to decay as: $\Zp \rightarrow t^{(\ast)} \bar{u}, \bar t^{(*)}u$. Events with, e.g., $u\bar u\to \Zp\Zp$ then lead to numerous like-sign top quark events, strongly constrained by data~\cite{Aaltonen:2008hx}. 

\begin{figure}[t]
\includegraphics[width=0.7\columnwidth]{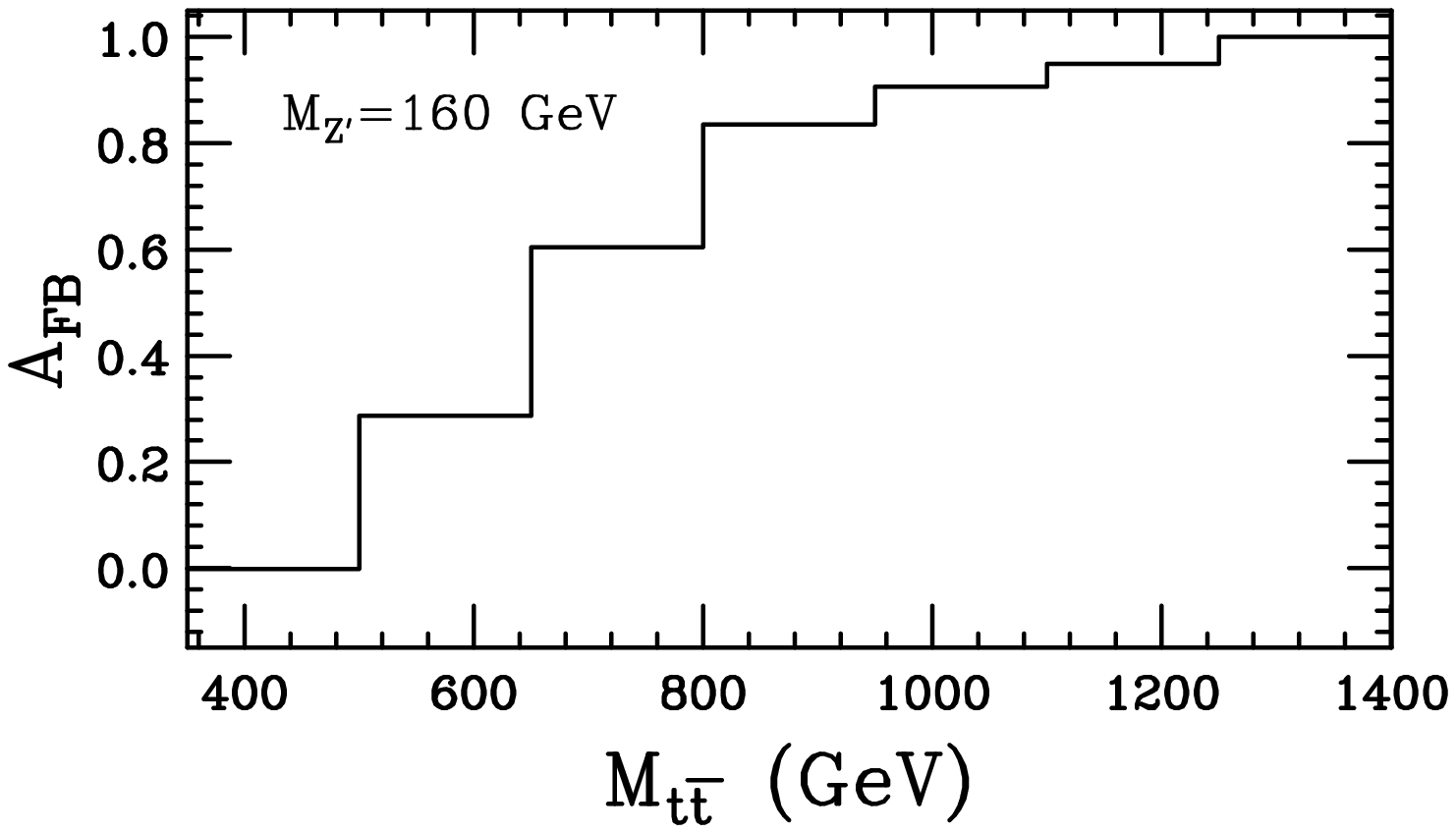}
\caption{$\Afb$ as a function of $\sqrt{\hat{s}}=\Mtt$ for $M_{\Zp} = 160$ GeV.}
\label{fig:rutherford}
\end{figure}

The model has three free parameters, $(g_{X}, \epsilon_{U}, M_{\Zp})$.  For $M_{\Zp} < m_{t}$ the phenomenology is essentially identical for all small $\epsilon_U \neq 0$.  This coupling is solely to provide the dominant two-body decay $\Zp\to u\bar u$.    We will show that a light $\Zp$, $M_{\Zp} \approx 160$ GeV with $\alpha_{X} \approx  2.4\times 10^{-2}$ is preferred when taking into account all considerations. We call this the ``best point" of the model.

Since we are giving non-trivial charges to the right-handed up-type quarks, 
bare Yukawa couplings are not invariant under $U(1)_{\Zp}$. We assume a Froggatt-Nielsen type mechanism \cite{Froggatt:1978nt} generates the Yukawa couplings. Chiral gauge anomalies can be satisfied, e.g., by adding two sets of extra heavy fermions of appropriate charge, and will not be discussed further here.

\textbf{Asymmetry and cross sections.}
The $t$-channel exchange of a new particle is a promising way to generate a large $\Afb$.  The cross-section in the forward, large $\Mtt=\sqrt{\hat s}$ region  is enhanced due to a Rutherford scattering peak.   We plot the asymmetry as a function of $\Mtt$ in~\Fig{fig:rutherford}, which shows this important effect.

A challenge for any model wishing to generate a large $\Afb$ is avoiding a too large modification of the $t \bar{t}$ production cross section.  The current measurement from 2.8 fb$^{-1}$ at CDF \cite{cdf:xtion-overall} is $\sigma(t \bar{t}) = 7.0 \pm 0.3$ (stat) $\pm 0.4$ (syst) $\pm 0.4$ (lumi) pb for $m_t = 175$GeV, in good agreement with the SM prediction of $\sigma(t \bar{t})_{SM} = 6.73-6.90$ pb \cite{Cacciari:2008zb,Kidonakis:2008mu,Moch:2008ai}, and is consistent with measurements from D0 \cite{dzero:xsection-total} that use smaller data sets.  

A typical color singlet $\Zp$ with flavor diagonal couplings does not interfere with the dominant (color-octet) QCD production process.  Thus, it is difficult to avoid a large shift of the $t\bar{t}$ production cross section as well as the appearance of a resonance. On the other hand, the $t$-channel exchange of our $\Zp$ in $p \bar{p} \rightarrow t \bar{t}$ interferes with QCD. It is possible then to have smaller modifications to the cross section while having a large contribution to $\Afb$. There is no resonance present in the $M_{t\bar{t}}$ spectrum.

We use MadGraph/MadEvent 4.4.17 \cite{Alwall:2007st} with CTEQ6.6M parton distribution functions~\cite{Nadolsky:2008zw} to generate event samples, and BRIDGE 2.0 \cite{Meade:2007js} to decay unstable particles.  We do not carry out parton showering or detailed detector simulation. We assume $m_t = 175$ GeV, and apply everywhere a QCD $K$-factor $K=1.31$ to match the SM prediction for $\sigma(t\bar{t})$, we fix renormalization and factorization scales at $\mu_R=\mu_F=m_t$.

\begin{figure}[t]
\includegraphics[width=\columnwidth]{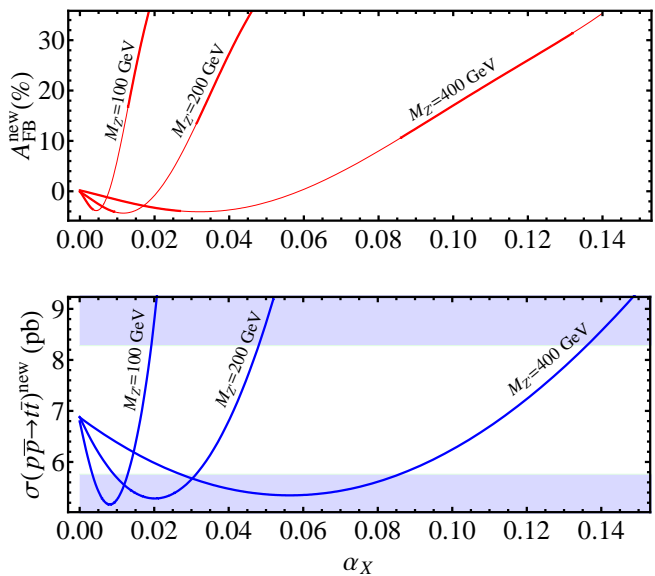}
\caption{$\alpha_X \equiv g_{X}^{2}/(4 \pi)$ versus $A_{FB}^{new}$ and $\sigma(t\bar{t})$ for $M_{\Zp}=100,200,400$ GeV (from the left).  In the lower panel, shaded regions deviate by more than $2\sigma$ from  $\sigma(t\bar{t})^{new}$.  Corresponding disfavored regions are shown as thinned lines in the upper plot. The superscript ``$new$'' emphasizes that only pure $\Zp$ and SM contributions are included (without fake processes).  These fakes leads to some subtlety in the allowed region, as discussed in the text.}
\label{fig:afbcx}
\end{figure}

We plot cross section and $A_{FB}^{new}$ in \Fig{fig:afbcx} as a function of $\alpha_{X}$ for three $\Zp$ masses.  $A_{FB}^{new}$ indicates the $\Afb$ induced only in the $t \bar{t}$ final state.  The SM NLO contribution (5\%) is not included. Similarly, the ``new'' in $\sigma(p\bar{p} \rightarrow t \bar{t})^{new}$ emphasizes that other (reducible) contributions that might enter the $t \bar{t}$ sample are not included.  They are discussed below.  

\begin{figure}[t]
\includegraphics[width=0.9\columnwidth]{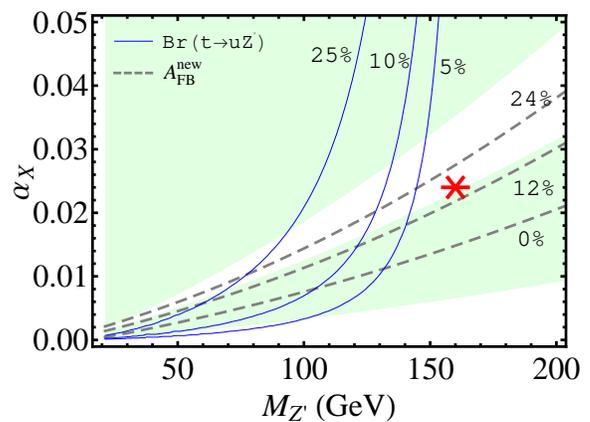}
\caption{A contour plot of $A_{FB}^{new}$ and BR($t \rightarrow \Zp u$) in the $\alpha_{X}$ - $M_{\Zp}$ plane. In colored regions, $\sigma(t \bar{t})^{new}$ deviates $2\sigma$ from of the measurement quoted in text. Parameter space around the red star is preferred.  A much larger $\alpha_{X}$ will gives too many like-sign top quarks, or a large distortion of the $\Mtt$ spectrum.  Larger masses lead to larger distortions of the $\Mtt$ spectrum, and smaller masses give a large branching ratio for $t \rightarrow \Zp u$, which leads to tension between measurement of top cross-sections in different channels.}
\label{fig:afbBR}
\end{figure}

Comparing the two panels of Fig.~\ref{fig:afbcx} indicates a potential simultaneous fit to a large $\Afb$ and the correct cross-section. However, new physics can contribute to final states that fake the $t \bar{t}$ final state.  This could pollute both the cross-section and the $\Afb$ measurement.  Reducible backgrounds that contaminate the sample arise, e.g., from $tt/\bar t\bar t$, $t\Zp/\bar t\Zp$ events, and modify the results of Fig.~\ref{fig:afbcx} by $\delta A_{FB}^{fake}, \delta\sigma(t\bar{t})^{fake}$. If $M_{\Zp} < m_t$, it is also important to include effects of exotic top decays $t \to u \Zp$ which can take events away from the registered $t\bar t$ cross-section. Assuming $\Zp$ decays are completely hadronic, they reduce the dilepton top cross section relative to the lepton+jets channel.   At CDF and D0, $t\bar{t}$ production is defined by specific final state topologies with at least one $b$ quark tag, several hard jets, and one (``$l+j$ sample") or two (``dilepton sample") charged leptons. CDF has measured $\sigma(t\bar{t}) = 7.2 \pm 0.75$ pb from the $l+j$ sample~\cite{cdf:xtion-lj}, and $6.7 \pm 0.98$ pb from the dilepton sample~\cite{cdf:xtion-ll}.  To avoid a too large discrepancy between these two channels, \Fig{fig:afbBR} shows that a light $\Zp (M_{\Zp} \lesssim 120$ GeV) is to be avoided.  For our ``best point'' we show comparisons with these cross sections in Table~\ref{tbl:best}.  Our simulation method is to construct event samples based on cuts detailed in~\cite{cdf:xtion-lj,cdf:xtion-ll}, and rescale the result by the inverse of the SM event selection efficiency (again using our simulation) to approximate their unfolding procedure.
 
For our best point (the red star in \Fig{fig:afbBR}), the total asymmetry is about $18\%$, see Table \ref{tbl:best}. This includes the SM $\alpha_s^3$ contribution, the $\Zp$ tree contribution, and contributions due to $\delta A_{FB}^{fake}$. The last is negative largely due to anti-correlation of $t$ direction with that of $u$ in $gu\to t\Zp$ production.  We estimate $|\delta A_{FB}^{fake}|$ at a few percent, not quite canceling with the $+5\%$ SM contribution. There is a small uncertainty in this estimate, as the kinematics of these events are not identical to those analyzed in the $t \bar{t}$ events.

Table \ref{tbl:best} shows the top quark asymmetry and the inferred $t\bar t$ cross section of our best point in the $l+j$ and dilepton channels. The asymmetry is high, and the cross-sections are within errors of the measurements. A prediction is the inferred cross-section from the dilepton sample should be less than from the $l+j$ sample: $t\Zp/\bar t\Zp$ events produce relatively more events in the $l+j$ sample than in the dilepton sample. In addition, events with exotic top decays ($t\rightarrow  \Zp u \rightarrow u \bar{u} u$) may contribute to the $l+j$ sample but not the dilepton sample.

\begin{table}[t] \begin{center} 
\begin{tabular}{cccc}
\hline
 & $l+j$ (pb)  & dilepton (pb) &  $A_{FB}^{tot}$ \% \\
\hline \hline
$M_X=160$ GeV, $\alpha_X=0.024$ &  $7.5$ & $5.8$ & $18$ \\
Measurements~\cite{cdf:xtion-lj,cdf:xtion-ll,cdf:afb} & $7.2\pm 0.8$ &  $6.7\pm 1.0$ & $19\pm 7$ \\
\hline
\end{tabular} \end{center} \caption{$t\bar t$ cross sections and total asymmetry for our best parameter point compared with measurements at CDF.  There are measurements from D0 as well that use less data, and thus have larger error bars \cite{dzero:xsection-total,dzero:xsection-dilepton}
} \label{tbl:best} \end{table}

\begin{figure}
\includegraphics[width=0.9\columnwidth]{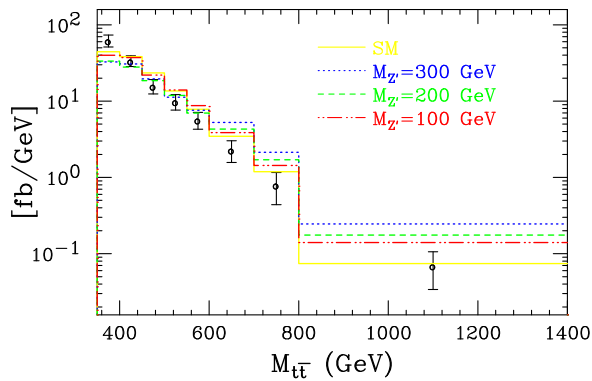}
\caption{The $\Mtt$ invariant mass spectrum. Data from the CDF measurement \cite{Aaltonen:2009iz} is shown along with our SM simulation. Also shown are $M_{\Zp} = 100,200,300$ GeV, with $\alpha_X = 0.013,0.03,0.055$, respectively.  Each $(\alpha_X, M_{\Zp})$ pair would provide an $A_{FB}^{new} \simeq 10\%$.}
\label{fig:mtt}
\end{figure}

\textbf{Additional collider constraints.}
Our model yields no resonances, but new $t$-channel physics modifies the $\Mtt$ distribution -- especially in the higher invariant mass bin due to the Rutherford enhancement.  This distribution has been measured by the CDF experiment in the lepton + jet channel ~\cite{Aaltonen:2009iz} and is  shown in Fig.~\ref{fig:mtt}.  We also show the apparent $\Mtt$ from this model, which includes contributions from fake processes. We observe that the heavier the $\Zp$, the more the last bin deviates from the measurement.  This is because the Rutherford singularity (beneficial to the generation of the $\Afb$) is most effective at $\Mtt \gg M_{\Zp}$.  A higher mass $\Zp$ will thus need higher $\alpha_{X}$ because it cannot take full advantage of the singularity, leading to larger distortion of $\Mtt$.  Thus, lighter $\Zp$ is favored.

The $t$-channel exchange of $\Zp$ can also produce like-sign top-quark events $uu(\bar u\bar u)\to tt(\bar t\bar t)$, which have been discussed in a different context by \cite{BarShalom:2008fq}. Like-sign tops can be observed as like-sign dilepton events plus $b$ tag(s). CDF has measured only 3 such events with 2 fb$^{-1}$ of data \cite{Aaltonen:2008hx}. The SM expectation is also small but with large error: $2.1 \pm 1.8$ events. Our best point model predicts 5--6 events. Higher $\Zp$ mass models produce too many such events from, e.g., $t\Zp\to tt+\bar u$ if $Z\to u\bar u$ (i.e., $\epsilon_U$) is not large enough. For very large $\epsilon_U$, constraints on the  $\Zp$ from the dijet channel \cite{Aaltonen:2008dn} become important.  This is another reason why we desire $M_{\Zp}<M_t$.  This combination of constraints largely determines the location of the ``best point'' of Fig.~\ref{fig:afbBR}.

There is another reason that $\Zp \rightarrow t^{(\ast)} \bar{u}$ decays are potentially dangerous.  CDF has measured the ratio of $t\bar{t}$+0 jets to $t\bar{t}+ n$ jets, with a result consistent with the SM value \cite{Aaltonen:2009iz}.  If the $\Zp  \rightarrow t^{(\ast)} \bar{u}$ decays are present, they will preferentially contribute to the $t\bar{t}$+n jets, potentially at a dangerous level.  A non-zero $\epsilon_{U}$ removes this conflict.

There are also potential contributions to the single-top sample.  As discussed earlier, with $\epsilon_U \neq 0$, decays of the $\Zp \rightarrow u \bar{u}$ dominate.  Then the dominant contribution to the single-top sample comes from the process $u g \rightarrow t \Zp \rightarrow t u \bar{u}$.  This process (after multiplication by a $K$-factor of 1.3), gives a production cross section of 3 pb.  This is comparable to the SM prediction for single-top production (2.9 pb).  The measurement of single-top at D0 and CDF \cite{Abazov:2009ii,Aaltonen:2009jj} relies on a multivariate analysis using detailed kinematic information to extract the single-top events from a large background dominated by $W+$heavy flavored jets.  These backgrounds are nearly an order of magnitude larger than the signal described here.  So, it is not possible to say without such a detailed experimental analysis whether a constraint presently exists.  As a test of this model, the Tevatron experiments might look in the single-top sample and see whether it is possible to discern a resonance in the two light-flavored jets corresponding to the $\Zp$.  This measurement might also be possible at the Large Hadron Collider (LHC).

\textbf{Flavor physics.}
One might wonder whether the novel flavor violation of this model might be constrained by $B$ meson decays.  The structure of the theory wherein off-diagonal couplings are limited to the right-handed up-type quarks make this model particularly safe.  

Box diagrams containing both intermediate $W$ and $\Zp$ bosons can communicate flavor violation to the $B$ sector, giving operators of the form ${\mathcal O}_{d,s} = (\bar{b}\Gamma d_{i})(\bar{u} \Gamma u)$, where $d_{i}= d,s$.   However, these operators are only 0.3\% (4\%) for $d_{i}= d (s)$ of the SM tree level CKM-suppressed contributions to similar operators, and are of no concern.   Moreover, even the CKM-suppressed ${\mathcal O_{s}}$ is negligible compared to the penguin contribution in processes like $B \rightarrow K \pi$, see, e.g.,  \cite{Gronau:2005gz}.  If present, a flavor-off diagonal coupling involving a charm quark could give a dangerous contribution to $D$--$\bar{D}$ mixing.  So, the charges of the right-handed quarks under the $\Zp$ must be such that any off-diagonal couplings to the charm are supressed. 

Flavor changing neutral currents of SM gauge bosons are also induced by one-loop penguin diagrams where $\Zp$ runs in the loop with one off-diagonal and one diagonal coupling. The $t \to u g$ measurement by CDF \cite{Aaltonen:2008qr} gives the strongest bound. For $(M_{\Zp}, \alpha_X)$ pair with $A_{FB}^{new} \simeq 10\%$, this measurement translates into a relatively weak bound $\epsilon_U \lesssim {\cal O}(1)$. 

\textbf{Structure of Couplings.}
As an existence proof, we note that we can reproduce the desired couplings by starting with $U(1)_X$ charges of the three right-handed up-type quarks of $\{ -1+\epsilon_U, \, 0+\epsilon_U, \, 1+\epsilon_U \}$.  To find the couplings in the mass basis, we perform the rotation on the right handed up quarks.  For appropriate Yukawa couplings, there exists a unitary matrix, $W_{u}^{R}$, that transforms the diagonal couplings above into the desired predominantly off-diagonal couplings.  The up-type Yukawa couplings are determined in terms of this $W_{u}^{R}$ and the $V_{u}^{L}$, which enters the CKM matrix $V_{CKM} \equiv V_{u}^{L} V_d^{L \dagger}$.  

A direction similar to the minimal $U(1)_{\Zp}$ discussed here is to introduce an $SU(2)_{flavor}$ gauge symmetry under which the $(t_{R}, u_{R})$ form a doublet.   The $\Afb$ can then be explained through the $t$-channel exchange of the $W'$ gauge bosons.  Because the $W'$ carries a conserved ``top-charge'', its production and exchange no longer contribute to like-sign top quark production.  Avoiding a large (negative) contribution to the $\Afb$ from, e.g., $ug \rightarrow W^{\prime} t$ requires the introduction of a small $W^{\prime}-\bar{u}-u$ coupling. This can be engineered if the $SU(2)_{flavor}$ is broken by multiple Higgs fields, for example a triplet and a doublet.  Searches for like-sign top will not be decisive in determining whether nature realizes this approach.  The other phenomenology may be quite similar to that presented here: differences between the lepton+jet  and the dilepton $\sigma_{t\bar{t}}$ cross-sections will still be present.  This model predicts and additional contribution to the single top sample as well.  

\textbf{Discussion.}
The exchange of a $t$-channel $\Zp$ with a $\Zp-u-t$ coupling can produce a large $\Afb$ consistent with  other top quark observables. Our best parameter point $M_{\Zp} = 160$ GeV with $\alpha_X=0.024$ generates $A_{FB}^{tot} \simeq 18\%$, about four times larger than the SM prediction. The most constraining collider observable is the search for like-sign top quarks events, which is ameliorated by the introduction of small flavor diagonal couplings. The diagonal couplings are essential also in $\Mtt$ distribution as well as for ensuring $\sigma(t\bar t$+0 jets$)$/$\sigma(t\bar t \, + \geq 0$ jets) is consistent with observation.   More precise measurements of the top cross-section and searches for like-sign tops at the Tevatron should be decisive for this model.  

Although heavier $\Zp$ $(m_{\Zp} > m_{t})$ suffers from a relatively large like-sign top signal and a disfavored $\Mtt$ distribution, narrow regions of the parameter space might remain.  In this region, one is pushed to a large $\epsilon_U \approx 0.3$ (larger values are constrained by dijet searches).   In this case, the maximum $A_{FB}^{tot} \lesssim 10 \%$. 

If the true asymmetry at the Tevatron is greater than 15\% and is caused by our $\Zp$ theory, the LHC will also have many opportunities to discover its effects. Certainly the most important effect is again the like-sign dilepton channel. Deviations are more likely to show up there in the early years of LHC running than through the top quark asymmetry (The LHC, being a $pp$ machine, must form the asymmetry with respect to the $t\bar t$ boost direction).


Finally, we comment that a new gauge boson is not the only $t$-channel approach to generating an asymmetry.
Scalar bosons in the $t$-channel may play a similar role as a vector boson. However, unlike vector boson, the intereference of the scalar with the SM diagram generates a negative asymmetry. The sign can be flipped if the scalar couples to anti-top rather than top: $\epsilon_{\alpha 
\beta\gamma}\phi^\alpha t^\beta u^\gamma$ 
A scalar which has electric charge $4/3$ and is a QCD fundamental 
could play this role.  The current version of MadGraph/MadEvent and CalcHEP cannot handle the color flow of the coupling 
\cite{Kang:2007ib}. We leave detailed exploration of this direction for future study.

\noindent
{\it Acknowledgments:} The authors would like to thank D.~Amidei for numerous discussions, P.~Ko, J.~Shao,  M.~Strassler, D.Whiteson, and members of CERN and KIAS for many useful comments. SJ is supported by Samsung Scholarship and DOE. AP is supported by NSF CAREER Grant NSF-PHY-0743315. AP and JDW are supported in part by DOE Grant DE-FG02-95ER40899. HM is supported in part by World Premier International Research Center Initiative (WPI Initiative), MEXT, Japan, in part by the U.S. DOE under Contract DE-AC03-76SF00098, and in part by the NSF under grant PHY-04-57315.
\bibstyle{apsper.bst}
\bibliography{Afb}

\end{document}